\documentclass[twocolumn,aps,prb,preprintnumbers,amsmath,amssymb,superscriptaddress]{revtex4}
\usepackage{graphicx}
\begin{document}

\title {Conduction-band tight-binding description for Si applied to P donors}
\author{A. S. Martins}
\affiliation{Instituto de F\'{\i}sica, Universidade Federal Fluminense, 
24210-340, Niteroi-RJ, Brazil }
\author{Timothy B. Boykin}
\affiliation{Departament of Electrical and Computer Engineering, The University of
Alabama in Huntsville, 
AL 35899
}
\author{Gerhard Klimeck}
\affiliation{Network for Computational Nanotechnology, Electrical and Computer
Engineering, Purdue University, 
IN 47907, and 
Jet Propulsion Laboratory, California Institute of Technology, 
CA 91109}
\author{Belita Koiller}
\affiliation{Instituto de F\'{\i}sica, Universidade Federal do Rio de Janeiro, Cx.P.
68.528, 21945-970, RJ, Brazil}
\date{\today}

\begin{abstract}
A tight-binding parametrization for silicon, optimized to correctly reproduce
effective masses as well as the reciprocal space positions of the conduction-band
minima, is presented. 
The reliability of the proposed parametrization is assessed by performing systematic
comparisons between the descriptions of donor impurities in Si using this
parametrization and previously reported ones.
The  spectral decomposition of the donor wavefunction demonstrates the importance of
incorporating full band effects  for a reliable representation, and that an
incomplete real space description results from a truncated reciprocal space
expansion as proposed within the effective mass theory.

\end{abstract}

\maketitle

Advances in semiconductor device fabrication, particularly Si-based devices, have
benefited from the progressive miniaturization and integration of their constituent
parts, including detailed control of the doping process.
The theoretical approach utilized for device modeling should be able to resolve
charge-density variations on an atomic scale, and in this respect tight-binding (TB)
models are particularly adequate, as they 
provide atomistic descriptions of structural and electronic properties of
solids.\cite{aldo03}
Since the pioneering work of Slater and Koster,\cite{slater} TB is conceived as an
empirical method, where  orbital energies and hoppings are parameters to be adjusted
in order to reproduce relevant properties in the band structure of the solid.
TB methods rely on the choice of a suitable parametrization of the Hamiltonian,
which includes the choice of a basis set, range of hopping coupling, etc. For the
group IV  and III-V semiconductors  
one might expect that the minimal $sp^3$ basis with first-neighbors coupling should
suffice to describe the essential physics.  
However, this model is not able to reproduce the gaps of the indirect gap materials
like Si and AlAs.\cite{chadi75} A simple scheme to correct this deficiency 
consists in introducing an excited s-like orbital, $s^{\ast}$, to improve the
description of the conduction band (CB).\cite{vogl} 
Another limitation of first-neighbor models based on $s$ and $p$ orbitals alone is 
that they predict an infinite value of the transverse effective mass at the $X$ point. 
This anomaly may be overcome either by adding the five $d$ orbitals in the
first-neighbor basis 
set,\cite{jancu,boykin2004}  or by inclusion of second-neighbor (2nn) interactions
in the $sp^3s^*$ model.\cite{boykin} 

Comparison between the different TB parametrizations for Si available in the
literature shows  that the one that best reproduces the relevant bulk material
properties is the one proposed by Klimeck {\it et al},\cite{klimeck} based on the
first and second-neighbors $sp^3s^*$ model.
The reciprocal space position of the CB minima, however, is not well fitted. 
Silicon is an indirect gap material, with 6 CB minima along the equivalent $\Delta$
lines, 
85\% of the way between $\Gamma$ and $X$. 
In Ref.~\onlinecite{klimeck}, a genetic algorithm (GA) fitting was adopted, and the
target position for the minima was considered as 75\% between $\Gamma$ and $X$. 
For some applications\cite{boykin2004,koiller01} 
it is known that a shift in the position of the CB minimum may lead to
unsatisfactory results. 
We present here an improved parametrization, also obtained through the GA
methodology,\cite{klimeck} reproducing the correct positions of the CB edges. 

GA-based optimization employs stochastic methods 
which do not require constrains on continuity of solution space. 
In the  case considered here, the TB parameters are varied in order to get an
optimal set 
that reproduces given material properties, denoted as target values. 
Details on the method are discussed in Ref.~\onlinecite{klimeck}: 
We refer to the parametrization published there as P075, and to the one proposed
here as P085. 
Both parametrizations give the $k$-space positions of the six band minima at six
equivalent points along $\Delta$ lines, with the position of the minimum at
$\Delta_{min}=0.75 (2\pi/{\rm a_{Si}})$ for P075 and $\Delta_{min}=0.85 (2\pi/{\rm
a_{Si}})$ for P085, where ${\rm a_{Si}} = 5.431$\AA~is the conventional cubic
lattice parameter for Si. 
Table \ref{tbl:tabpar} presents the TB parameters for Si for both parametrizations.
The main difference between them consists in allowing non-zero 2nn hoppings
$V_{ss^{\ast}}(110)$ 
and $V_{s^{\ast}s^{\ast}}(110)$ in the optimization set for P085.  
All 2nn hoppings are consistently smaller than the first-neighbors.
The 2nn hopping parameters were determined to adjust the finer details of the target
properties, and were not constrained to have values or signs expected from physical
considerations.\cite{slater} 

\begin{table}
\caption{Parameters for the TB models (in eV). 
\label{tbl:tabpar}}

\begin{tabular}{l|c|c}
\hline \hline
Parameter & $\rm P075$ & $\rm P085$ \\ \hline
$E_s(000)$ & -4.81341 & -4.848054  \cr
$E_p(000)$ &  1.77563 & 1.787118  \cr
$E_{s^{\ast}}(000)$ &  6.61342 & 5.608014  \cr
$V_{ss}\left(\frac{1}{2}\frac{1}{2}\frac{1}{2} \right)$ & -8.33255 & -8.259704  \cr
$V_{xx}\left(\frac{1}{2}\frac{1}{2}\frac{1}{2} \right)$ &  1.69916 & 1.697556  \cr
$V_{xy}\left(\frac{1}{2}\frac{1}{2}\frac{1}{2} \right)$ &  5.29091 & 5.351079  \cr
$V_{sp}\left(\frac{1}{2}\frac{1}{2}\frac{1}{2} \right)$ &  5.86140 & 5.822197  \cr
$V_{s^{\ast}p}\left(\frac{1}{2}\frac{1}{2}\frac{1}{2}\right)$ & 4.88308 & 4.864480\cr
$\lambda_{SO}$  &  0.04503 & 0.014905  \cr
$V_{ss}(110)$  &  0.01591  & 0.029958  \cr
$V_{s^{\ast}s^{\ast}}(110)$  & 0.00000  &  0.191517  \cr
$V_{ss^{\ast}}(110)$  & 0.00000 & 0.007036  \cr
$V_{sx}(110)$  &  0.08002 & 0.161749  \cr
$V_{sx}(011)$  &  1.31699 & 0.885988  \cr
$V_{s^{\ast}x}(110)$  & -0.00579  & -0.095653  \cr
$V_{s^{\ast}x}(011)$  &  0.50103  & 0.966257  \cr
$V_{xx}(110)$  &   0.00762  &  0.037296  \cr
$V_{xx}(011)$  &  -0.10662  & -0.132810  \cr
$V_{xy}(110)$  &   0.55067  &  0.619876  \cr
$V_{xy}(011)$  &  -2.27784  & -2.496288  \\ \hline \hline
\end{tabular}
\end{table}

The input and the calculated properties  
are presented in Table \ref{tbl:tabcomp}, where the first column shows the material
properties to be represented by the TB model and the second column shows the
corresponding experimental values.\cite{sidata} 
These constitute the input targets in the GA code.
The remaining columns give the calculated properties and the respective deviations
from the experimental targets for P075 and P085, as well as for a recently proposed
first-neighbor $sp^3d^5s^{\ast}$ parametrization, which we denote by
P1nn.\cite{boykin2004}
\begin{table}
\caption{Optimization targets and optimized material properties for the P075, P085
and P1nn models.
Except for  $\Delta_{min}$, which is specific for different models, the target values 
correspond to experimental data given in Ref.~\onlinecite{sidata}.
\label{tbl:tabcomp}}
\begin{tabular}{l|r|rr|rr|rr}
\hline \hline
Property & Target & P075 & $\%dev$ & P085 & $\%dev$ & P1nn & $\%dev$ \\ \hline
$\Delta_{min}$& 0.750 & 0.758 & 1.067 & ~ & ~ & ~ & ~ \cr
$~$            & 0.850 & ~ & ~ & 0.8480 &-0.235 & 0.813 & -4.35 \cr\hline
$E_c^\Gamma$ & 3.350 & 3.353 & 0.089 & 3.350 & -0.013 & 3.399 & 1.44 \cr
$E_{c}^{\Delta_{min}}$ & 1.130 & 1.129 &-0.050 & 1.130 & -0.042 & 1.131 & 0.09 \cr
$m_{Xl}^{\ast}$ & 0.916 & 0.916 &-0.030 & 0.916 & 0.050 & 0.891 & -2.73 \cr
$m_{Xt}^{\ast}$ & 0.191 & 0.191 & 0.020 & 0.191 & 0.007 & 0.201 & -5.23 \cr
$m_{lh}^{\ast}[001]$ & -0.204 & -0.198 & 3.082 & -0.204 &-0.060 &-0.214 & -4.90 \cr
$m_{lh}^{\ast}[011]$ & -0.147 & -0.146 & 0.525 & -0.148 &-0.568 &-0.152 & -3.40 \cr
$m_{lh}^{\ast}[111]$ & -0.139 & -0.139 & 0.395 & -0.140 &-0.610 &-0.144 & -3.60 \cr
$m_{hh}^{\ast}[001]$ & -0.275 & -0.285 &-3.643 & -0.277 &-0.786 &-0.276 & -0.36 \cr
$m_{hh}^{\ast}[011]$ & -0.579 & -0.581 &-0.338 & -0.574 & 0.869 &-0.581 & -0.34 \cr
$m_{hh}^{\ast}[111]$ & -0.738 & -0.737 & 0.119 & -0.727 & 1.466 &-0.734 & 0.54 \cr
$m_{so}^{\ast}$ & -0.234 & -0.237 & -1.487 & -0.239 &-2.162 &-0.246 & -5.13 \cr
$\Delta_{so}$ & 0.015 & 0.145 & -0.067 & 0.015 & 0.030 & 0.016 & 4.90 \\ \hline \hline
\end{tabular}
\end{table}
We note from Table \ref{tbl:tabcomp} that the P075 and P085 parametrizations give
consistently better agreement with the target values than the P1nn parametrization.
This statement cannot be generalized as a fundamental advantage of the
second-neighbor  
$sp^3s^*$  model over the nearest-neighbor $sp^3d^5s^*$ model.  
The P1nn  set\cite{boykin2004} satisfies additional requirements beyond bulk
behavior properties.  
On-going work indicates that the $sp^3d^5s^*$ model can be fit to match the Si bulk
properties just as well as the second-neighbor $sp^3s^*$ model.  The primary
advantage of the nearest-neighbor $sp^3d^5s^*$ model is its straightforward
incorporation of strain distortions.\cite{boykin2002} 
In the donor description below, no strain distortions are considered, and the
second-neighbor model P085 provides an unprecedented representation of the Si CB. 

The reliability of TB $sp^3s^*$ second-neighbors parametrizations has been recently 
verified by studies of shallow donors in GaAs\cite{martins02} and in
Si.\cite{martins04} 
We perform the same kind of study here, 
and discuss the effect of the CB minimum position in different aspects of the donor
problem. 
We write the Hamiltonian for the impurity problem as\cite{form1}
$H=\sum\limits_{ij}\sum_{\mu \nu}h_{ij}^{\mu\nu}c_{i\mu}^{\dagger}c_{j\nu}+\sum_
{i,\nu} U(R_i)c_{i\nu}^{\dagger}c_{i\nu}$ 
where $i$ and $j$ label the atomic sites, $\mu$ and $\nu$ denote the atomic orbitals
and $R_{i}$ is the distance between  site $i$ and the impurity site. 
The impurity potential is taken as a screened Coulomb potential,
$U(R_i)=-e^2/(\varepsilon R_i)$  ($\varepsilon = 12.1$ for Si).  
At the impurity site 
it is assigned the value $U(R_{i}=0)=-U_{0}$, a parameter describing central cell
effects characteristic of the substitutional species, and taken here as an
adjustable parameter.  We do not include spin-orbit corrections in the present
calculations.

The eigenstates of $H$ are determined for a system where a single impurity is placed
in a cubic supercell containing $N=8L^{3}$ atoms in the diamond structure, where $L$
is the length of the supercell edge in units of  
$\rm a_{Si}$. 
We adopt periodic boundary conditions, and 
large supercells\cite{martins02} (up to $10^6$ atoms) were treated within a
variational scheme\cite{araujo,zunger1994} where the ground state wavefunction and binding
energy $E_{L}$ are obtained by minimizing 
$\left\langle\Psi\left|\left(H-\varepsilon_{ref}\right)^{2} \right|\Psi
\right\rangle$. 
The reference energy  $\varepsilon_{ref}$ is chosen within the gap, nearest to the
CB edge.

The eigenfunctions of $H$ in the basis of atomic-like orbitals 
are written as $|\Psi_{TB} (\mathbf{r})\rangle = \sum_{i\nu} a_{i\nu}|\phi_{\nu}
(\mathbf{r}-\mathbf{R}_i)\rangle $, where the expansion coefficients $a_{i\nu}$ give
the probability amplitude of finding the electron in the orbital $\nu$ at site
$\mathbf{R}_i$. 
The overall charge distribution is conveniently described through the TB envelope
function squared,\cite{tania} 
\begin{equation}
|\Psi_{EF}({\mathbf{R}_i})|^2 = \sum_{\nu} |a_{i\nu}|^2.
\label{envfunc} 
\end{equation}

\begin{figure}
\includegraphics[width=70 mm]{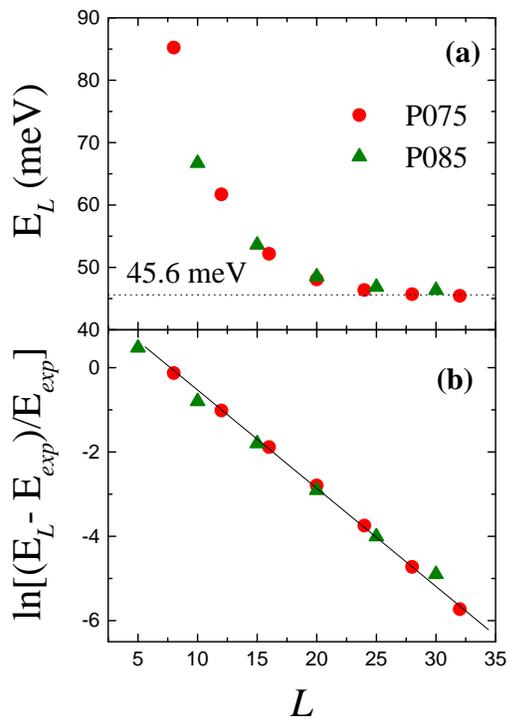}
\caption{ (Color online) (a) Convergence of the donor ground state binding energy
towards the experimental value (dotted line gives $E_{exp}=45.6$~meV) with the
supercell size $L$, for   
$U_0=U_P$, namely 1.48 eV and 1.26 eV for P075 and P085 respectively. 
(b) Same data plotted as $\ln[(E_L-E_{exp})/E_{exp}]$ vs $L$.
The linear behavior (the line is a best fit for all data points) 
indicates that the convergence of the binding energy with the supercell size $L$
follows the same exponential law for both parametrizations.
\label{fig1}}
\end{figure}
\begin{figure}
\includegraphics[width=70 mm]{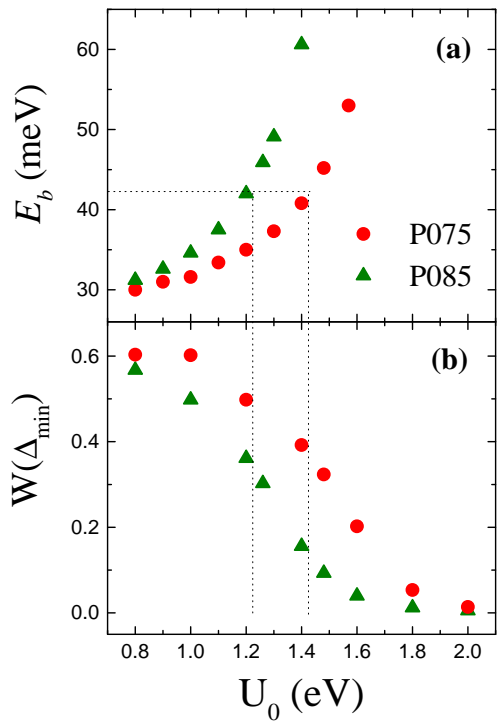}
\caption{(Color online) (a) Binding energy of the ground state as a function 
of the on-site perturbation strength $U_0$. 
The dotted lines indicate the value $U_0 = U_P$ that reproduces the experimental
Si:P $A_1$ state binding energy. 
(b) Total spectral weight at the CB edges for the ground impurity state.
\label{fig2}}
\end{figure}
\begin{figure}
\includegraphics[width=70 mm]{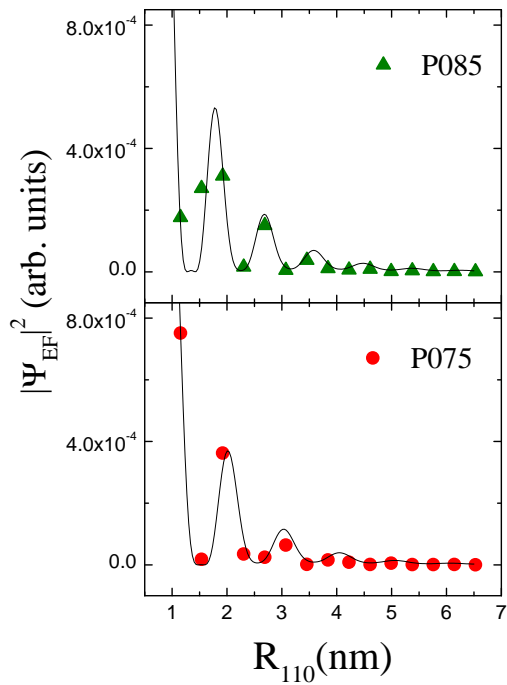}
\caption{(Color online) TB envelope function squared for the donor ground state
along the [110] direction.
The lines are the corresponding effective mass $|\psi|^2$ results. 
\label{fig3}}
\end{figure}

In Fig.~\ref{fig1}, a convergence study of the donor ground state binding energy as
a function of the supercell size $L$ is presented for P075 and P085. 
One can observe in Fig.~\ref{fig1}(a) that for supercell sizes $L>25$ the calculated
binding energies reproduce the experimental value ($E_{exp}=45.6$ meV) taking
$U_{0}=1.48$ and $U_{0}=1.26$ eV for P075 and P085, respectively.   
We denote these values by $U_P$, as they are determined by tuning the on-site
potential $-U_0$ in order to give the  converged value of $E_b$ in agreement with
experiment for P donors in Si. Fig.~\ref{fig1}(b) presents a plot of
$\ln[(E_L-E_{exp})/E_{exp}]$ vs $L$. The linear behavior obtained here indicates
{{\it the same} exponential convergence of $E_L$ to the experimental value for both
parametrizations: 
$E_L \sim E_{exp} + \tilde{E} e^{-L/\lambda}$. 

We determine the first excited state energy\cite{martins04} by varying the value of
$\varepsilon_{ref}$ in 
$\left\langle\Psi\left|\left(H-\varepsilon_{ref}\right)^{2} \right|\Psi \right\rangle$.
Keeping $U_0=1.26$ eV, we obtain the binding energy of the first excited state to be
30.2 meV, in good agreement with  the P075 result, 32.4 meV. 

The behavior of the binding energy with $U_0$ is presented in Fig.~\ref{fig2}(a),
where the dotted lines indicate  the value of $U_P$ for each parametrization leading
to $E_b = E_{exp}$. As noted in Ref.~\onlinecite{martins04}, in the weak
perturbation limit 
the binding energies converge to the effective mass theory (EMT) prediction in its
simplest formulation~\cite{emtsi} (single-valley approximation), $\sim 30$ meV. It
is interesting that this behavior is the same for both parametrizations. As $U_0$
increases, P085 tends to give higher binding energies than P075, resulting in a
smaller value of $U_P$ for P085. 

One can also characterize the donor ground state by its orbital averaged spectral
weight\cite{tania} at ${\bf k}=\Delta_{min}$, 
$W(\Delta_{min})=\frac{2}{N}\sum_{\mu=1}^6\sum\limits_{ij\nu}e^{i{\bf{k}_\mu}
\cdot({\bf R}_{i}-{\bf R}_{j})} a_{i\nu}a_{j\nu}$, 
plotted in Fig. \ref{fig2}(b) as a function of $U_0$. 
The EMT approach presumes that $W(\Delta_{min})\sim 1$, 
allowing the donor state to be well described in a basis of Bloch states at the
CB-edge $\bf k$ points. 
However, one can notice in Fig. \ref{fig2}~(b) that this is not the case: Even for
the smallest values of $U_0$ the spectral weights at $\Delta_{min}$ are well below
saturation (one) for both parametrizations, implying that an incomplete description
may result from EMT in this case. 
The spectral weight at $U_P$ is 0.32 for P075 and 0.30 for P085.
These relatively low spectral weights indicate that multiple {\bf k} points, other
than those corresponding to the six CB-edges of Si, contribute to the donor
wavefunction expansion within any reciprocal-space based approach. 

Within single-valley EMT 
the ground state for donors in Si is six-fold degenerate.\cite{emtsi} 
Valley-orbit interactions\cite{baldereschi} lead to a non-degenerate ground state
wavefunction of $A_1$ symmetry,
\begin{equation}
\psi ({\bf r}) = \frac{1}{\sqrt{6}}\sum_{\mu = 1}^6 F_{\mu} ({\bf r})
u_\mu({\bf r})e^{i {\bf k}_{\mu}\cdot{\bf r}}\,,
\label{eq:sim}
\end{equation}
where $\phi_\mu({\bf r}) = u_\mu({\bf r})e^{i {\bf k}_{\mu}\cdot{\bf r}}$ 
are the pertinent Bloch wavefunctions, and the envelope functions are given by 
$ F_{z} ({\bf r}) = (1/\sqrt{\pi a^2 b})\ e^{-[(x^2+y^2)/a^2 + 
z^2/b^2]^{1/2}}$ for $\mu=z$ and equivalently for the other $\mu$ values.
The effective Bohr radii for Si are $a=2.51$~nm 
and $b=1.44 $~nm.\cite{koiller01} 
Fig.~\ref{fig3} presents a comparison between the TB envelope function calculated
from (\ref{envfunc}) along the $[110]$ direction, for both P075 and P085
parametrizations (data points) with the corresponding EMT results obtained from
(\ref{eq:sim}) (solid lines), with $a$ and $b$ are given above, but with a different
normalization to conciliate the TB and EMT wavefunctions on the same scale.
The values of ${\bf k}_{\mu}$ used in (\ref{envfunc}) are consistent with the
respective reciprocal space location of the CB minima. Note that the oscillatory
behavior due to interference among the plane-wave parts of the six $\phi_\mu$ is
well captured by the TB envelope function.
The period of the oscillations is different for P075 and P085, as given by the
corresponding wavevectors. 

Good agreement between TB and EMT is 
restricted to distances from the impurity site larger than $\sim$~1~nm. 
This means that at large distances  
the EMT expansion of the donor wavefunction in only six {\bf k} points, as given in
Eq.(\ref{eq:sim}), is capable of reproducing its main features (except for
normalization, of course). Closer to the impurity, particularly at the impurity site, the TB results become much larger than the EMT prediction.

The wavefunctions obtained from P085 and P075 agree reasonably well in the central
cell region. One way to quantify this agreement is through the probability to find the donor
electron inside of a sphere of radius $R_c$: $Q(R_c)=\sum_{R \leq R_c}|a_{i\nu}(R)|^2$. 
Taking for $R_c$ the 2nn distance, the ratio of $Q(R_c)$ obtained from 
the two parametrizations is $Q_{\rm P085}(R_c)/Q_{\rm P075}({R_c})=1.15$.

In summary, we find good agreement between the results obtained within P085 and P075
for (i) the exponential convergence law for the ground state binding energy with
supercell size, (ii) the binding energy of the first excited state, (iii) the
spectral weight of the ground state wavefunction at $\Delta_{min}$, (iv) the
probability that the donor electron is within  the central cell up to the impurity's
2nn. Both parametrizations also capture the donor wavefunction oscillations predicted
within EMT; the main difference regards the period of the oscillations.
In applications where the quantitative aspects of the oscillatory behavior of the
wavefunction is important, the P085 parametrization is thus capable of providing a
better description. The importance to represent the P impurity wavefunction in a full band, atomistic representation is demonstrated through its spectral decomposition,  
and by direct comparison between the TB results and EMT.

{\bf Acknowledgments}~~
We thank F.J. Ribeiro for fruitful discussions. 
BK thanks the hospitality of CMTC at the University of Maryland.
This work was partially supported in Brazil by FAPERJ, CAPES, FUJB, 
Instituto do Mil\^enio de Nanoci\^encias/MCT, and  
at the Jet Propulsion Laboratory, Caltech by NASA.
Funding for GK was provided by JPL, NASA-ESTO, ARDA, ONR, and NSF. 

\bibliography{martins_resub}
\end{document}